\documentclass[runningheads]{svmult}
\usepackage{makeidx}   
\usepackage{graphicx}  
\usepackage{subeqnar}  
\usepackage{multicol}  
\usepackage{cropmark} 
\usepackage{physprbb}  
\def\Journal#1#2#3#4{{#1} {\bf #2}, #3 (#4) } 

\def\NPB{{\em Nucl. Phys.} B}
\def\PLB{{\em Phys. Lett.}  B}
\def\PRL{\em Phys. Rev. Lett.}
\def\PRD{{\em Phys. Rev.} D}
\def\ZPC{{\em Z. Phys.} C}

\def\PRL{Phys. Rev. Lett.}

\input colordvi

\def\r2{\sqrt 2}
\def\beq{\begin{equation}}
\def\eeq{\end{equation}}
\def\beqn{\begin{eqnarray}}
\def\eeqn{\end{eqnarray}}

\def\sinW2{\sin^2\theta_W}

\def\mz2{M_{z}^2}
\def\c2b{\cos 2\beta}

\def\mz{M_z}

\def\Fq2{F_{2}(q^2)}

\def\beq{\begin{equation}}
\def\eeq{\end{equation}}
\def\amu{a_\mu}

\def\tanbeta{{\rm tan}\beta}
\def\gmin2{(g-2)_\mu}

\def\sec2w{sec^2\theta_W}

\def\r2{\sqrt 2}
\def\beq{\begin{equation}}
\def\eeq{\end{equation}}
\def\beqn{\begin{eqnarray}}
\def\eeqn{\end{eqnarray}}

\def\sinW2{\sin^2\theta_W}

\def\mz2{M_{z}^2}
\def\c2b{\cos 2\beta}

\def\mz{M_z}

\def\Fq2{F_{2}(q^2)}
\def\sq2{\sqrt{2}}

\def\sec2w{sec^2\theta_W}

\begin{document}
\title*{Status of Supersymmetry in the Light of Recent Experiments}
\toctitle{Status of Supersymmetry in the Light of Recent Experiments}
%
%
\titlerunning{Status of Supersymmetry}
%

\author{Utpal Chattopadhyay\inst{1}
\and Achille Corsetti\inst{2}
\and Pran Nath\inst{2}
  }
\authorrunning{Chattopadhyay, Corsetti, Nath}
%
%
\institute{Harish-Chandra Research Institute, Chhatnag Road, 
Jhusi, Allahabad 211019, India
\and Department of Physics, Northeastern University, Boston,
MA 02115, USA}

\maketitle              

\begin{abstract}
In this talk we discuss the constraints on supersymmetry arising from data
from a number of recent experiments. There appears to be good cumulative
evidence from experiment in favor of positivity of the sign of
the Higgs mixing parameter $\mu$. Implications of this result for
Yukawa unification are discussed since Yukawa unification is sensitive
to the $\mu$ sign. An analysis of dark matter under the constraints of
Yukawa unification is also given. It is shown that the simultaneous
imposition of all existing constraints sharply defines the parameter
space of models. Specifically models with nonuniversality of gaugino masses
provide a simple resolution to the positivity of the $\mu$ parameter and 
Yukawa unification. Implications of these results for colliders and for
the next generation of dark matter searches are also discussed.
\end{abstract}

\section{Introduction}
Over the recent past accumulation of data from experiments that
probe physics beyond the standard model has begun to constrain 
new physics. In this talk we discuss the constraints  on
supersymmetry arising from these experiments. If 'recent 
experiments' is interpreted to include the experimental results over
the past decade then there is an impressive body of data which taken
together can put important limits on the parameter space of 
supersymmetric models. Perhaps at the top of the list here is 
the precision data on the gauge coupling constants\cite{pdg} 
which looks very favorable for supersymmetry in that the 
minimal supersymmetric standard model with a low lying sparticle
spectrum is in excellent accord with the data within 
$1\sigma$-$2\sigma$ (for a review see\cite{couplinguni}).
A small discrepancy that might exist can be easily taken account of
with the help of Planck scale corrections\cite{hall}. However, the 
gauge coupling unification does not put very stringent limits 
on the sparticle masses or on the sign of the Higgs mixing parameter
$\mu$. Constraints on the sparticle masses and on the
$\mu$ parameter arise from the experiment on the 
flavor changing neutral current 
process $b\rightarrow s+\gamma$\cite{cleo,belle,aleph}.
Here one finds that the sparticle spectrum is constrained
and one sign of $\mu$ is preferred 
(the positive sign in the standard convention\cite{sugra}) and for
the other $\mu$ sign most of the parameter space of a class of
supersymmetric models is eliminated\cite{susybsgamma1,susybsgamma2}.
 Interestingly the central value 
of the $b\rightarrow s+\gamma$ experiment\cite{cleo,belle,aleph} 
is somewhat lower than the central value of its 
standard model prediction\cite{smbsgamma} providing a slight hint of
 a supersymmetric
contribution. This is so because a supersymmetric loop correction from the 
chargino exchange can provide a contribution with a negative sign.  
Another important constraint comes from the recent lower limit
from the LEP experiment on the Higgs mass\cite{lephiggs}. 
Thus before the LEP experiment closed down, it was  able to place a lower 
limit on the Higgs mass of $115$ GeV for the Standard Model. For the 
supersymmetric case the  lower limit is a function of $\tan\beta$ and
can be as low as about 90 GeV for large $\tan\beta$. Again these
lower limits impose important constraints on the parameter space of
supersymmetric models. 
Additional constraints on supersymmetric 
models emerge from the $g-2$ Brookhaven experiment\cite{brown}. 
We will discuss the $g-2$ experiment and its implications in detail
in Sec.2. One of the main results that emerges from the BNL data
is the positivity of the $\mu$ parameter which is in accord with
the sign preferred by the $b\rightarrow s+\gamma$ constraint.
Further, the positivity of $\mu$ has important implications
for Yukawa coupling unification. This topic will be discussed 
in detail in Sec.3. 
One of the important features of supersymmetry is that it 
provides a candidate for cold dark matter under the assumption of
R parity conservation. Thus using renormalization group 
a class of supergravity models lead naturally to the lightest 
neutralino to be the lightest supersymmetric particle (LSP) and
hence a candidate for dark matter\cite{lsp}.
Further, this class of models can produce just the right amount of dark
matter that is indicated by the current astrophysical
observations. The price one pays for generating the right amount of 
dark matter is to further constrain the parameter space of
supersymmetric models. This topic
will be discussed in Sec.4. Finally, we note that the 
SuperKamiokande experiment has now reached its maximum 
sensitivity\cite{superk}
for the detection of the mode $p\rightarrow \bar \nu K^+$ preferred
 by supersymmetric models and has placed a new limit of 
 $\tau(p\rightarrow \bar \nu K^+)>1.9\times 10^{33}yr$. This
 result puts the minimal supersymmetric grand unified models
 under considerable stress. We will briefly discuss
 this topic in the conclusion.

\section{$g-2$ Experiment}
The anomalous magnetic moment $a=(g-2)/2$ is an important probe of
new physics beyond the standard model. Typically new physics contributions
to the anomalous moment obey
$a_l$(new physics)$\sim m_l^2/\Lambda^2$ where $\Lambda$ is
the scale of the new physics. Because of this $a_{\mu}$ is a much more
 sensitive probe of new physics than $a_e$. 
Last February the BNL experiment gave a new determination of $a_{\mu}$
with a substantially lower error\cite{brown} than previous 
measurements.
The current evaluation is $a_{\mu}^{exp}=11659203(15)\times 10^{-10}$.
As of a few months ago the Standard Model prediction consisting of
the qed, electro-weak and hadronic corrections amounted to
$a_{\mu}^{SM}= 11659159.7(6.7)\times 10^{-10}$\cite{czar1}. This gives
$a_{\mu}^{exp}- a_{\mu}^{SM}=43(16)\times 10^{-10}$ which is
a $2.6\sigma$ difference between theory and experiment.
However, over the past few months there has been a reevaluation of
the Standard Model prediction.
This revision arises from a change in sign
of the light by light hadronic correction. Thus previous analyses gave
for $a_{\mu}^{had}(LbL)$ the value\cite{hayakawa,bijnens}  
$a_{\mu}^{had}(LbL)=-8.5(2.5)\times 10^{-10}$. 
More recent 
evaluations\cite{knecht,hkrevised}
however, find the sign of $a_{\mu}^{had}(LbL)$ to be opposite to 
that of the previous determinations. Thus the analysis of 
Knetch {\it et. al} gives \cite{knecht} $a_{\mu}^{had}(LbL)=8.3(1.2)\times 10^{-10}$
and a follow up analysis by  
Hayakawa and Kinoshita gives\cite{hkrevised} 
$a_{\mu}^{had}(LbL)=8.9(1.5)\times 10^{-10}$. Support for the change
in sign also comes from partial analysis by Bijnens {\it et.al}\cite{Bijnens2}
 which gives $a_{\mu}^{had}(LbL)=8.3(3.2)\times 10^{-10}$ and by 
Blockland {\it et.al.}\cite{Blokland}
which gives $a_{\mu}^{had}(LbL)=5.6\times 10^{-10}$.
Taking the average of the first two 
and correcting the standard model prediction one gets 
$a_{\mu}^{SM}= 11659176.8(6.7)\times 10^{-10}$. This leads to the 
$1.6\sigma$ deviation between experiment and the standard
model result, i.e., 
\begin{equation}
a_{\mu}^{exp}-a_{\mu}^{SM}= 26(16)\times 10^{-10} 
\end{equation}
We note in passing that there is no unanimity yet on the size 
of the light by light hadronic correction. Indeed, very recently
another evaluation of $a_{\mu}^{had}(LbL)$ based on chiral perturbation
theory was given in Ref.\cite{Ramsey}. This analysis
finds $a_{\mu}^{had}(LbL) =(5.5^{+5}_{-6}+3.1\hat C)\times 10^{-10}$ 
where $\hat C$ stands for corrections arising from the subleading contributions.
The result of Eq.(1) corresponds to a value of $\hat C\sim 1$. However, 
the authors of  Ref.\cite{Ramsey} indicate that a $\hat C$ range of  
$-3$ to 3 or even larger is not unreasonable. In addition, to the 
uncertainties associated with the light by light contribution, one
also has the errors associated with the evaluation of the
$\alpha^2$ correction to the vacuum polarization. In the evaluation
of Eq.(1) we used the $\alpha^2$ correction to the vacuum polarization
as given by the analysis of Ref.\cite{davier} which gives
$a_{\mu}^{had}(\alpha^2 vac.pol.) =(692.4 \pm 6.2)\times 10^{-10}$. However, 
there are several other evaluations of this quantity\cite{hadronic}
and this subject is still a topic of further investigation.  

We discuss next the  SUSY contribution to $a_\mu$. 
The first analysis of the supersymmetric 
correction to $a_\mu$ was given soon after the development of 
SUGRA unified models\cite{msugra}.
At the one loop level one has $\amu^{SUSY}$$=a_{\mu}^{\tilde \chi^{\pm}}$
$+a_{\mu}^{\tilde \chi_i^0}$.  
For the CP conserving case one has that 
the chargino contribution is the larger one. It is given by\cite{yuan} 
\beq
 a_{\mu}^{\tilde \chi^{\pm}}={{m^2_\mu} \over {48{\pi}^2}}
  {{ {A_R^{(a)}}^2} \over
{m_{\tilde{\chi}_a^\pm}^2}}F_1(\left({{m_{\tilde \nu}} 
\over {m_{\tilde \chi^{\pm}_a}}}\right)^2)+
{{m_\mu} \over{8{\pi}^2}} {{A_R^{(a)}
A_L^{(a)}} \over {m_{\tilde \chi^\pm_a}}} 
F_2(\left({{m_{\tilde \nu}} \over {m_{\tilde \chi^{\pm}_a}}}\right)^2)
\eeq
where 
$A_L(A_R)$ are the left(right) chiral amplitudes. Now the chiral 
interference term proportional to $A_LA_R$, particularly the contribution from
the lighter chargino term typically dominates the
chargino exchange contribution. The above amplitude
contains some very interesting properties\cite{lopez,chatto}.
 One finds that typically
$ A_L\sim 1/\cos\beta$ and  because of this $\amu^{SUSY}\sim \tan\beta$.
Further, it is easy to show that $A_L$ depends on the sign of $\mu\tilde m_2$
 and thus effectively
the sign of $\amu^{SUSY}$ is controlled by the sign of $\mu\tilde m_2$
(where we use the sign covention of Ref.\cite{sugra}).
Thus one finds that typically\cite{lopez,chatto} 
\beq
\amu^{SUSY}>0, \tilde m_2\mu>0; ~~\amu^{SUSY}<0, \tilde m_2\mu<0
\eeq
\begin{figure}[h]
\begin{center}
\includegraphics[width=.5\textwidth]{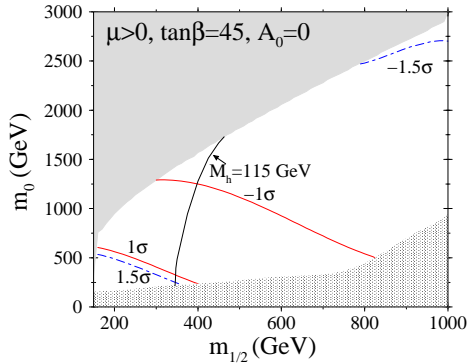}
\end{center}
\caption[]{Regions corresponding to 
 the $1.5\sigma$ and the $1\sigma$ constraints of $a_{\mu}^{SUSY}$ 
for $\tan\beta=45$ from Ref.\cite{ccnyuk}. The top left gray regions do not 
satisfy the radiative electroweak symmetry breaking requirement or the lighter
chargino mass limit, whereas
 the bottom 
patterned regions are typically discarded by stau becoming the LSP.
The bottom patterned region near the 
higher $m_{1/2}$ side and on the border of the white allowed regions are 
discarded because of CP-odd Higgs boson turning tachyonic at the tree level.}
\label{eps1.2}
\end{figure}
\begin{figure}[h]
\begin{center}
\includegraphics[width=.5\textwidth]{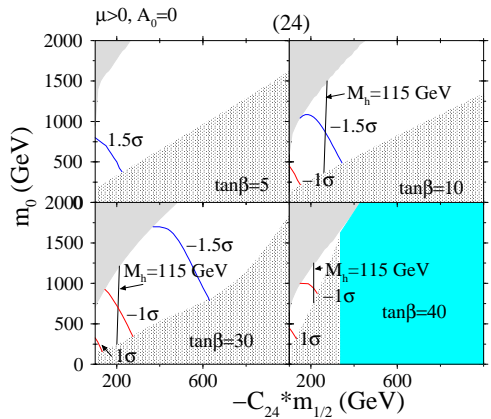}
\end{center}
\caption[]{Allowed regions  corresponding to  $1.5\sigma$ and  
$1\sigma$  constraints $a_\mu^{SUSY}$ for nonuniversal gaugino mass
 scenario of the SU(5) 24 plet case from Ref.\cite{ccnyuk}.
The top gray regions correspond to disallowed areas via radiative 
electroweak symmetry breaking constraint. 
The bottom patterned regions for  
$\tan\beta=$~5,10 and 30 are typically eliminated via the 
stability requirement of the Higgs potential at the GUT scale.
Part of the region 
with large $|c_{24} m_{1/2}|$ and large $m_0$ bordering the 
allowed (white) region for $\tanbeta=30$ is eliminated via the limitation of 
the CP-odd Higgs boson mass turning tachyonic at the tree level.  
For $\tanbeta=40$ most of the region 
(patterned and shaded) to the right of 
the allowed white small region is 
eliminated because of $\lambda_b$ going to 
the non-perturbative domain due to a large supersymmetric 
correction to the b quark mass.}
\label{eps1.1}
\end{figure} 
\begin{figure}[h]
\begin{center}
\includegraphics[width=.5\textwidth]{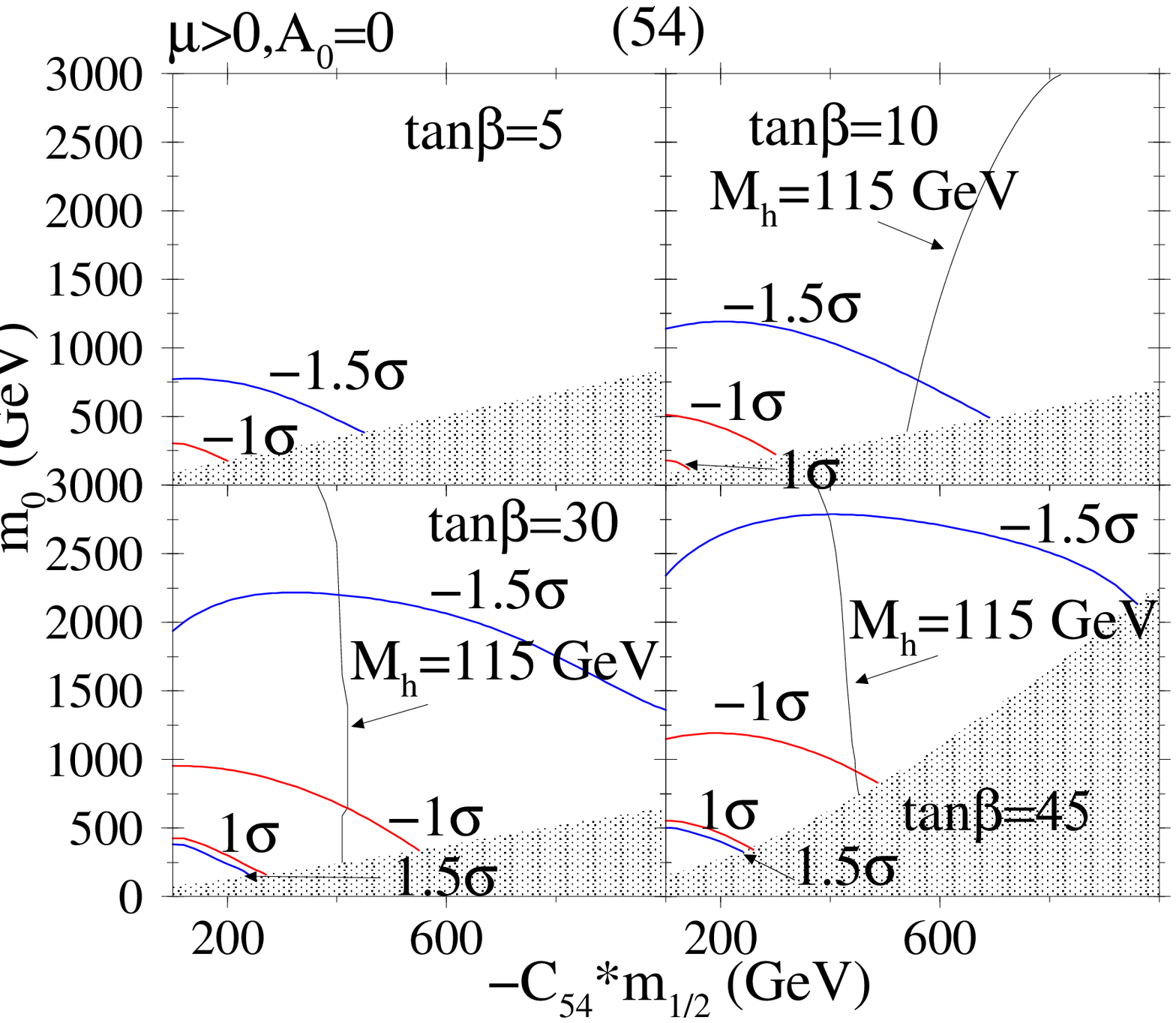}
\end{center}
\caption[]{ Allowed regions corresponding to $1.5\sigma$ and 
$1\sigma$ constraints on $a_\mu^{SUSY}$ for nonuniversal gaugino
mass scenario of SU(10) 54 plet case from Ref.\cite{ccnyuk}. 
Here the nonuniversal Higgs scalar 
parameters are given by $m_{H_1}^2=1.5m_0^2$ and $m_{H_2}^2=0.5m_0^2$.                   
The regions with patterns 
are discarded for reasons similar to as in Fig.2.
  }
\label{eps1.2}
\end{figure} 
More recently the absolute signs of the supersymmetric contribution
was checked by taking the supersymmetric limit and it was shown 
that, as expected, in this limit  $a_{\mu}(SM)+a_{\mu}(SUSY)=0$\cite{in}.
Soon after the BNL result of Ref.\cite{brown} was announced, several
analyzes on its implications for the sparticle spectrum were 
carried out\cite{chatto2,everett} using the result
$a_{\mu}^{exp}-a_{\mu}^{SM}=(43\pm 16)\times 10^{-10}$ which found 
upper limits on sparticle masses well within reach of the LHC.
 Specifically in the work of 
Ref.\cite{chatto2} using  a 2$\sigma$ error corridor around the
announced difference $a_{\mu}^{exp}-a_{\mu}^{SM}$ 
so that $ 10.6\times 10^{-10}< a_{\mu}^{SUSY}<76.2\times 10^{-10}$
it is found that the BNL data implies that in mSUGRA 
the following upper bounds on sparticle  masses hold\cite{chatto2}
$m_{\tilde \chi^{\pm}}\leq 650 GeV,$ $ m_{\tilde \nu}$ $\leq 1.5 TeV$$
 $$~(\tan\beta\leq 55)$, 
$m_{1/2}\leq 800 GeV,$ $m_0\leq 1.5 TeV$$ ~(\tan\beta\leq 55)$. 
In view of 
the revised difference of Eq.(1) a new analysis of the allowed 
parameter space in mSUGRA was carried out in Ref.\cite{ccnyuk}.
Results are displayed in Fig.1 with $1\sigma$ and $1.5\sigma$ 
error corridors. One finds that the $1\sigma$ upper limits with
the revised result are exactly the same as the $2\sigma$ upper
limits of the previous analysis\cite{chatto2}. Specifically, in this
case the squarks and gluinos upper limits lie below 2 TeV. 
Now the LHC can explore squarks/gluinos up to 2 TeV\cite{cms}. 
Thus the revised $1\sigma$ upper limits gotten from the 
$g-2$ result imply that sparticles should become visible at the LHC.
The upper limits for the $1.5\sigma$ case, however, are significantly
larger as may be seen from Fig.1 and one finds that part of the allowed
parameter space lies outside the reach of the LHC.  
An interesting aspect of the BNL data is that it determines the 
sign of $\mu$ given the sign of $\tilde m_2$. Thus assuming CP
conservation one finds that the 
BNL data determines the $sign(\mu\tilde m_2)$ to be positive.
Thus $\mu$ is determined to be positive for a wide class of
supersymmetric models where $\tilde m_2$ is positive.
Now $\mu$ positive is favored by
the $b\rightarrow s+\gamma$ constraint\cite{susybsgamma1} and
 a positive $\mu$ is also 
favorable for the satisfaction of the relic density constraints.
However, a positive $\mu$ is typically not favored by Yukawa coupling
unification. We will discuss this issue in detail in Sec.3. 

An interesting question is the possible effect of 
extra dimensions on the analysis of $g-2$ and whether such
contributions can be large enough to produce a significant
background for the supersymmetric effects. This question has been
examined in a variety of extra dimension models\cite{ny1,graesser}.
 The simplest
possibility considered is a model with one extra dimension
with the fifth dimension compactified\cite{ny1}. Specifically we consider 
a five dimensional model with the large extra dimension compactified
on $S^1/Z_2$ with radius R ($M_R=1/R=O(TeV)$).
The spectrum of this theory contains massless modes with N=1 SUSY in 4D, 
which precisely form the spectrum of MSSM in 4D. In addition one has
massive Kaluza-Klein modes which fall in N=2 multiplets. One of the
interesting things that happen in the above scenario is that the  mechanism
that generates corrections to $g-2$ also generates corrections to 
the Fermi constant\cite{ny1}. 
Now the
Standard Model prediction on $G_F^{SM}$ is in fairly good accord with
the experimental value. Thus the contribution arising from extra 
dimensions must be accommodated within the error corridor 
between the experimental value of $G_F$ and its standard model prediction.
 This constrains the size of the extra dimension so that $M_R\geq 3$ TeV. 
Now large extra dimensions affect the value of $a_{\mu}$ 
from contributions via the excitations of the $W,Z, \gamma$.
However, because of the fact that $M_R\geq 3$ TeV one finds that 
the effect of the extra dimension is numerically small, i.e., 
one finds that the corrections of the extra dimensions to $a_{\mu}$
is up to two or more orders of magnitude smaller than the supersymmetric
correction. Similar results hold for the case of strong gravity extra
 dimension models\cite{graesser} when one includes the constraints on extra 
dimensions from the current experiment\cite{gravity}.
Thus for practical purposes one finds that the contribution from 
extra dimensions does not produce a strong background to the supersymmetric
contribution. The above analysis also shows that $g-2$ is not a
sensitive probe of
extra dimensions. Perhaps the most effective way to probe extra 
dimensions is via  direct production of the Kaluza-Klein states at
colliders\cite{antoniadis,nyy}, where Kaluza-Klein masses up to 6 TeV can be
probed. Another important phenomenon is the effect of CP violation
of $g-2$. This was investigated in Refs\cite{in,icn}. and it was found that 
$g-2$ is a sensitive function of the phases. Indeed the constraints on 
$g-2$ can be utilized to constrain the phases themselves.
Further, if the new physics effect in the BNL experiment turns out to be
of size $\sim 10^{-9}$ then this would also be encouraging for the
possible observation of the muon electric dipole moment (EDM) 
$d_{\mu}$\cite{inmuedm}. 
This is so because the 
muon anomaly is related to the real part and the EDM to the imaginary
part of the same diagrams exchanging charginos and neutralinos.
These results are very interesting in view of a recent proposal to  probe 
$d_{\mu}$ with a sensitivity which is six orders of 
magnitude better than the current sensitivity\cite{muedm}.
\section{$b-\tau$ Unification and $\mu$ Sign}
$b-\tau$ unification prefers $\mu<0$\cite{bagger,deboer}. This issue is tied
closely with SUSY correction to the b quark mass. 
\beq
m_b(M_Z)=\lambda_b(M_Z)\frac{v}{\sqrt 2}\cos\beta(1+\Delta_b) 
\eeq
where $\Delta_b$ is the loop correction to $m_b$. 
The largest contributions to $\Delta_b$ arise from the gluino
 and the chargino exchanges\cite{susybtmass}
\beq
\Delta_b^{\tilde g}= \frac{2\alpha_3\mu M_{\tilde g}}
{3\pi}   \tan\beta I(m_{\tilde b_1}^2, m_{\tilde b_2}^2,M_{\tilde g}^2),~
\Delta_b^{\tilde \chi^+}= \frac{Y_t\mu A_t}
{4\pi} \tan\beta I(m_{\tilde t_1}^2, m_{\tilde t_2}^2,\mu^2)
\eeq
where $Y_t=\lambda_t^2/4\pi$ and $I(a,b,c) >0$. 
A useful criterion for $b-\tau$ unification and more generally for
Yukawa unification are the parameters 
$\delta_{ij}=|\lambda_i-\lambda_j|/\lambda_{ij}$ where
$\lambda_{ij}=(\lambda_i+\lambda_j)/2$.  
It is well known that $b-\tau$ unification requires a negative
contribution to the b quark mass. 
However, from Eq.(5) one finds that the dominant gluino exchange
contribution to $\Delta_b$ is  positive for a positive $\mu$ which
is not what is preferred by the $b-\tau$ unification constraint.
 How to reconcile the positivity of $\mu$  with
 the Yukawa unification has been discussed in several recent 
 works\cite{bf,bdr,ky,cnbtau}.
 We discuss here in some detail the scenarios
  where such a phenomenon arises naturally from
  gaugino mass nonuniversality. The basic mechanism here is rather
  simple. With nonuniversalities 
   one can arrange the gaugino masses  $\tilde m_2$ and 
 $m_{\tilde g}$  to have opposite signs which allows for  
 consistency with $g-2$ and $b-\tau$ unification with a positive
 $\mu$ sign. Further such opposite sign correlations between
 $\tilde m_2$ and  $m_{\tilde g}$ arise naturally for certain
 group structures. We discuss below SU(5) and SO(10) examples 
 where the above phenomenon manifests\cite{cnbtau}.  
  For the case of SU(5) unification the gaugino mass  matrix 
  in general transforms like the symmetric product of two adjoint
  representations of SU(5) and this product can be expanded as the
  sum of the representations 1,24, 75 and 200. Of these one finds
  that the 24 plet component yields  opposite signs for the 
  SU(2) and SU(3) gauginos. Thus one has\cite{anderson}  
 \beqn
 (24\times 24)_{sym}=1+24+75+200\nonumber\\
 M_3:M_2:M_1 = 2:-3:-1, ~~ 24- plet
 \eeqn
 A similar situation arises for the SO(10) case.
 Here the gaugino mass matrix transforms like the symmetric product of 
 two adjoint representations of SO(10) and this product can be expanded
 as the sum of the representations 1,54, 210 and 770. 
 Of these one finds that the 54 plet component yields opposite signs for the 
  SU(2) and SU(3) gauginos. In fact in this case there are more than
  one possible ways to achieve the opposite signs for the SU(2) and 
  SU(3) gauginos depending on the pattern of symmetry breaking.
  Thus one has\cite{chamoun}
\beqn
(45\times 45)_{sym}=1+54+210+770\nonumber\\
M_3:M_2:M_1 = 1:-3/2:-1 ~~54-plet\nonumber\\
M_3:M_2:M_1 = 1:-7/3:1 ~~54'-plet
\eeqn 
where the $54$-plet and the $54'$-plet correspond to two different
patterns of symmetry breaking.  In general the gaugino masses could
be a linear combination of the  different representations such that
$\tilde m_i(M_G)=m_{1/2}\sum_r C_r n_i^r$ where $n_i^r$ are 
characteristic of the representation r and $C_r$ are the relative
weights. In the analysis below we will consider only 
one representation 
in the sum. Thus for SU(5) we will consider the gaugino masses 
arising from the 24 plet representation and for the SO(10) case
we will consider the cases where the gaugino masses arise either
from the $54$ plet representation or from a $54'$ representation.
More generally, of course, one may have linear combinations of different
representations and the analysis given here can be extended easily to
these more general cases.

We consider first the 24 plet case of SU(5). Here  
an analysis similar to that of Fig.1 is given in Fig.2 where
the regions of the parameter space allowed by the $1\sigma$ and by
the $1.5\sigma$ $a_{\mu}$ constraint is exhibited\cite{ccnyuk}. We note that the  
allowed regions are significantly modified by the presence of 
nonuniversalities. An interesting phenomenon here is that 
 $\tan\beta$ in this case does not get much beyond the value 
$\tan\beta =40$ as the b quark Yukawa coupling gets into a  
nonperturbative domain because of large  supersymmetric corrections
to the b quark mass. 
A similar analysis for the 54 case of SO(10) is given in Fig.3
where we give an analysis of the allowed parameter space for 
$1\sigma$ and for the $1.5\sigma$ $a_{\mu}$ constrains. 
As is well known for the large $\tan\beta$ case of SO(10) one needs
nonuniversalities of the Higgs scalar mass parameters 
at the GUT scale to accomplish
radiative breaking of the electroweak symmetry. In the analysis of
Fig.3 we used $m_{H_1}^2=1.5m_0^2$ and $m_{H_2}^2=0.5m_0^2$.
The GUT boundary conditions for the above soft parameters were  varied 
(up to 50\%) to test the stability of the results. It was found that
there was no significant change in the results as a consequence of these
modifications.
The above scenarios produce a negative contribution to the b quark
mass for $\mu$ positive and thus lead to Yukawa unification for a positive
$\mu$ consistent with the $g-2$ and the  $b\rightarrow s+\gamma$
constraints.

\section{Implications for Dark Matter} 
Supersymmetric dark matter and its detection has been a topic
of considerable theoretical investigation over the years
(see Ref.\cite{ellisdark} for an overview of dark matter and
 Ref.\cite{detection} for a sample of recent works on the analysis
 of supersymmetric dark matter).
However, there are very few works on the study
of supersymmetric dark matter coupled with Yukawa unification\cite{gomez}.
Here we investigate the implications of the positivity of the $\mu$
sign and Yukawa unification for the detection rates in the 
direct detection of neutralino dark matter\cite{ccnyuk}. 
 We focus on the scalar cross section
\beq
\sigma_{\chi p}(scalar)=\frac{4\mu_r^2}{\pi}
 (\sum_{i=u,d,s}f_i^pC_i+\frac{2}{27}(1-\sum_{i=u,d,s}f_i^p)
 \sum_{a=c,b,t}C_a)^2 
\eeq
where  $f_i^p$ (i=u,d,s quarks) are the quark densities defined by
$ m_pf_i^p=<p|m_{qi}\bar q_iq_i|p>$. 
There are significant uncertainties associated with  the 
determination of $f_i^{p,n}$. An analytical solution for 
$f_i^{p,n}$ can be gotten by using\cite{cndark1}
  $\sigma_{\pi N}$, x, and $\xi$ as inputs
where $ <p|2^{-1}(m_u+m_d)(\bar uu+\bar dd|p>=\sigma_{\pi N}$,
$x=\sigma_0/\sigma_{\pi N}= 
<p|\bar uu+\bar dd-2\bar ss|p>/<p|\bar uu+\bar  dd|p>$, and
 $\xi= <p|\bar uu-\bar dd|p>/<p|\bar uu+\bar  dd|p>$.
In terms of the above variables 
 $f_i^{p}$  are  given by  
$f_u^p=(1+\xi)m_u\sigma_{\pi N}/m_p(m_u+m_d)$, 
$f_d^p=(1-\xi)m_d\sigma_{\pi N}/m_p(m_u+m_d)$, and 
similar relations hold for $f_i^{n}$. 
One also finds that $f_ip$ and $f_i^n$ satisfy the relation
$f_u^pf_d^p=f_u^nf_d^n$. 
We use the above to give a numerical evaluation of $f_i^{p,n}$ and  
its uncertainties. Using the most recent evaluation of $\sigma_{\pi N}$ of
$\sigma_{\pi N}=(64\pm 9)$GeV given by the 
SAID pion-nucleon database\cite{said}, ~~$x=0.55\pm 0.12$, 
the value of $\xi$ given by the 
 baryon mass splittings, i.e., by the formula  
$ \xi=(\Xi^-+\Xi^0-\Sigma^+-\Sigma^-)x/
(\Xi^-+\Xi^0+\Sigma^++\Sigma^--2m_p-2m_n) $ and using 
$\xi=   0.196 x$ one finds
$\xi=0.108\pm 0.024$. Additionally using the quark mass ratios
$m_u/m_d=0.553\pm 0.043, ~~m_s/m_d=18.9\pm 0.8$ 
one finds\cite{ccnyuk} $f_u^p=0.027\pm 0.005$, $f_d^p=0.038\pm 0.006$,
$f_u^n=0.022\pm 0.004$, $f_d^n=0.049\pm 0.007$, and 
$f_s^n=f_s^p=0.37\pm 0.11$. We use these quark densities to discuss the
implications of the scenarios discussed in Sec.3 for dark matter.
\begin{figure}[hbt]
\begin{center}
\includegraphics[width=.5\textwidth]{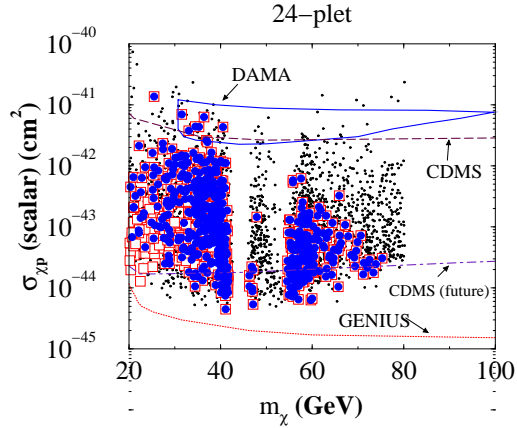}
\end{center}
\caption[]{Plot of the neutralino-proton scalar 
cross section $\sigma_{\chi p}$ vs the lightest neutralino 
mass $m_{\chi}$ for the SU(5) 24 plet case with the range of the 
parameters given in Fig.2 satisfying all the desired constraints including
the $b-\tau$ unification constraint so that $\delta_{b\tau}\leq 0.3$
from Ref.\cite{ccnyuk}.
The small crosses 
satisfy the $g_{\mu}-2$ constraints, the (blue) filled squares 
additionally satisfy the $b \rightarrow s+ \gamma$ limits and the (red) filled 
ovals satisfy all the constraints, i.e.,  the $g_{\mu}-2$ constraint,
the $b \rightarrow s+ \gamma$ constraint, and $\delta_{b\tau}\leq 0.3$.
The area enclosed by solid lines is excluded by the DAMA 
experiment\cite{dama}, the dashed line is the lower limit from the
CDMS experiment\cite{cdms}, the dot-dashed line is the lower limit 
achievable by CDMS in the future\cite{cdms} and the dotted line is the 
lower limit expected from the proposed GENIUS experiment\cite{genius}.}
\label{eps1.3}
\end{figure}

An interesting issue concerns the question if the parameter space
consistent with the constraints discussed in Sec.3 will pass the 
test of neutralino relic density constraints for cold dark matter (CDM). 
Current estimates show that CDM satisfies the constraint
 $0.02\leq \Omega_{\chi} h^2\leq 0.3$ where $\Omega_\chi$ is the ratio
 of the CDM relic density and the critical relic density needed to
 close the universe, and $h$ is the hubble parameter in units of 
 100 km/sMpc. Additionally, we will require that for SU(5) we satisfy
 the $b-\tau$ unification constraint of $\delta_{b\tau}\leq 0.3$ and
 for SO(10) the $b-t-\tau$ unification\cite{shafi} constraint of
 $\delta_{ij}\leq 0.3$ where $\{i,j\}=b,t,\tau$.  
 \begin{figure}[hbt]

\begin{center}
\includegraphics[width=.5\textwidth]{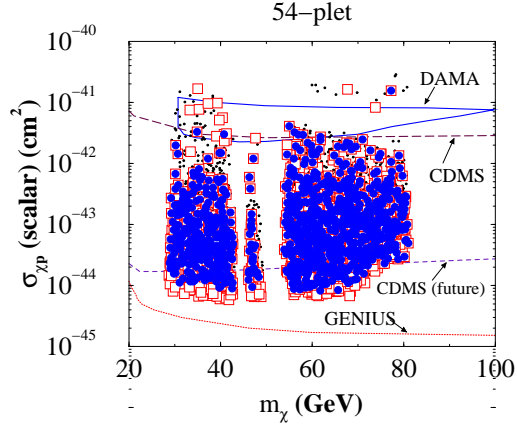}
\end{center}
\caption[]{Plot of the neutralino-proton scalar 
cross section $\sigma_{\chi p}$ vs the neutralino mass $m_{\chi}$ for the 
SO(10) 54 plet case  from Ref.\cite{ccnyuk}
satisfying all the desired constraints including
the $b-\tau$, $b-t$ and  $t-\tau$ unification constraint so that 
$\delta_{b\tau}, \delta_{bt}, \delta_{t\tau}\leq 0.3$. The small crosses 
satisfy the $g_{\mu}-2$ constraints, the (blue) filled squares 
additionally satisfy the $b \rightarrow s+ \gamma$ limits and the (red) filled 
ovals satisfy all the constraints, i.e.,  the $g_{\mu}-2$ constraint,
the $b \rightarrow s+ \gamma$ constraint, and 
Yukawa unifications with $\delta_{b\tau},\delta_{bt} \delta_{t\tau} 
\leq 0.3$.}
\label{eps1.2}
\end{figure}
\begin{center}
\begin{tabular}{|c|c|c|c|}
\multicolumn{4}{c}{Table 1: ~Sparticle masses for 24, 54, 
 $54'$ cases from Ref.\cite{ccnyuk} } \\
\hline
   & {\bf 24} (GeV) & {\bf 54} (GeV) & {\bf $54'$} (GeV)\\
\hline
  $~\chi_1^0$~ & ~~~~32.3 - 75.2~~~~ &~~~~32.3 - 81.0~~~~ & 32.3 - 33.4\\
 \hline
  $\chi_2^0$ & 96.7 - 422.5 & 94.7 - 240.8 & 145.7 - 153.9\\
 \hline
  $\chi_3^0$ & 110.5 - 564.3 & 301.5 - 757.1 & 420.9 - 633.8\\
 \hline
  $\chi_4^0$ & 259.2 - 575.9 & 311.5 - 759.7 & 427.6 - 636.9\\
 \hline
  $\chi_1^\pm$ & 86.9 - 422.6 & 94.6 - 240.8 & 145.8 - 153.9\\
 \hline
  $\chi_2^\pm$ & 259.9 - 577.2 & 315.1 - 761.6 & 430.7 - 639.2\\
 \hline
  $\tilde g$ & 479.5 - 1077.2 & 232.5 - 580.3 & 229.8 - 237.4 \\
 \hline
  $\tilde\mu_1$ & 299.7 - 1295.9 & 480.5 - 1536.8 & 813.1 - 1196.3\\
 \hline
  $\tilde\mu_2$ & 355.1 - 1309.3 & 489.8 - 1482.7 & 835.3 - 1237.6\\
 \hline
  $\tilde\tau_1$ & 203.5 - 1045.1 & 294.2 - 1172.6 & 579.4 - 863.7\\
 \hline
  $\tilde\tau_2$ & 349.6 - 1180.9 & 422.6 - 1311.7 & 704.6 - 1018.3\\
 \hline
  $\tilde u_1$ & 533.6 - 1407.2 & 566.7 - 1506.4 & 822.9 - 1199.8\\
 \hline
  $\tilde u_2$ & 561.1 - 1443.0 & 584.7 - 1544.6 & 849.6 - 1232.6\\
 \hline
  $\tilde d_1$ & 535.1 - 1407.5 & 580.3 - 1546.2 & 845.1 - 1232.5\\
 \hline
  $\tilde d_2$ & 566.7 - 1445.2 & 590.1 - 1546.7 & 853.3 - 1235.2\\
 \hline
  $\tilde t_1$ & 369.9 - 975.2 & 271.5 - 999.6 & 513.7 - 819.9\\
 \hline
  $\tilde t_2$ & 513.7 - 1167.6 & 429.4 - 1107.4 & 599.4 - 848.2\\
 \hline
  $\tilde b_1$ & 488.2 - 1152.8 & 158.1 - 1042.0 & 453.2 - 749.9\\
 \hline
  $\tilde b_2$ & 532.3 - 1207.0 & 396.6 - 1159.2 & 610.5 - 880.4\\
 \hline
  $h$ & 104.3 - 114.3 & 103.8 - 113.3 & 108.1 - 110.9\\
 \hline
  $H$ & 111.9 - 798.8 & 151.5 - 1227.6 & 473.4 - 831.9\\
 \hline
  $A$ & 110.5 - 798.8 & 151.4 - 1227.6 & 473.4 - 831.9\\
 \hline
  $\mu$ & 96.0 - 559.5 & 291.1 - 752.7 & 413.1 - 628.4\\
 \hline
 \end{tabular}\\
\end{center}

\noindent
  Assuming that CDM is composed entirely of
 neutralinos one finds that the parameter space allowed by the
 constraints discussed in Sec.3 indeed allow for the satisfaction of 
 the relic density constraints. The sparticle spectrum including the
 relic density constraints is discussed in Table~1. 
 The spectrum exhibited in Table~1 is consistent with the
 criterion of naturalness (see, e.g., Ref.\cite{ccn}) and it would
 be interesting to discuss the possible signal such as the trileptonic
 signal\cite{trilep} that emerge from this spectrum. 
 One may also discuss the 
 detection rates for the direct detection of dark matter in these
 scenarios. In Fig.4 an analysis is given of the maximum and the
 minimum scalar cross section  $\sigma_{\chi -p}$ for the 24 plet case
 of SU(5) as a function of
 the the neutralino mass. The region of the parameter space
 where is signal is claimed by DAMA\cite{dama} is exhibited in Fig.4. Also
 exhibited in Fig.4 are the  experimental upper limits from the CDMS
 experiment\cite{cdms} and the limits that the CDMS experiment
  and the proposed
 GENIUS experiment\cite{genius} will be able to achieve in the future.
 The analysis of Fig.4 shows that the future CDMS experimental limits
 will probe a major part of the parameter space of the 24 plet model,
 while the GENIUS detector will probe the entire parameter space 
 of the  24 model. A similar analysis for the 54 plet case of SO(10) 
 is given in Fig.5 with the same conclusions as for the case of Fig.4.
 A analysis for the $54'$ case is given in Ref.\cite{ccnyuk}.

\section{Conclusion}
We summarize now our conclusions. First if the
  $a_{\mu}^{exp}-a_{\mu}^{SM}$ difference persists at a perceptible 
 level, i.e., at the level  $\sim 10^{-9}$ or larger, then 
 direct observation of new physics is implied.  Assuming new
 physics is SUSY, we expect that most of the sparticles such as 
 $\tilde g, \tilde q, \tilde \chi^{\pm}, \chi^0$ etc should become visible
  at the LHC. In this paper we also discussed the implications  
  of a positive $\mu$ implied by the BNL data. Now it is well known that
  a positive $\mu$ is preferred by the $b\rightarrow s+\gamma$ constraint
  in that a large part of the parameter space is allowed by this 
  constraint for a positive $\mu$. 
  Further, a positive $\mu$ is beneficial for the direct search for 
  dark matter since a large part
  of the parameter space is available for the satisfaction of the relic
  density constraints. One downside to a positive $\mu$ is that
  Yukawa unification is more difficult. 
  However, it appears possible to overcome this problem.  
   One way to accomplish this is to use nonuniversalities of the gaugino 
  masses and the analysis here shows that a simultaneous 
satisfaction of all the constraints, i.e., the $g-2$, $b\rightarrow s+\gamma$ 
and $b-\tau$ 
unification constraints is possible.
 Models of the type discussed here  which satisfy these  constraints
 typically  produce a low lying Higgs boson mass which 
can be probed at RUNII of the Tevatron. Further, the sparticle spectrum
of these models is typically also low lying and can be fully probed at 
the LHC.  Thus these models 
 can  also be fully tested via direct detection of dark matter since a 
detector such as GENIUS can probe the entire parameter 
space of these models.
 Finally, we wish to draw attention to proton decay in supersymmetric
 GUT models\cite{pdecay}. 
 The current data from 
 SuperKamiokande indicates that the minimal SUSY GUT models 
including the minimal SU(5) and SO(10) models
 are under stress\cite{dmr,mp}. 
The above situation arises in part due to an improved value
 of $\beta_p$\cite{aoki} (the coefficient of the three quark operator between
 p and the vacuum state) and  due to an increase in the lower limits
 on the proton decay lifetime\cite{superk}. 
 The current experimental constraint
 on the muon anomaly further tends to destabilize the  proton.
 Several approaches to correct the situation have been proposed recently
 which  mostly involve dealing with non-minimal 
 models\cite{altarelli,ns}.

~\\
This work was  supported in part by NSF grant PHY-9901057.\\



\begin{thebibliography}{1}


\bibitem{pdg}
Eur. Phys. J. {\bf C15}, 1(2000). 



\bibitem{couplinguni}
For a review  see, K.~R.~Dienes,
Phys.\ Rept.\  {\bf 287}, 447 (1997).


\bibitem{hall}
L.~J.~Hall and U.~Sarid,
Phys.\ Rev.\ Lett.\  {\bf 70}, 2673 (1993);
T.~Dasgupta, P.~Mamales and P.~Nath,
Phys.\ Rev.\ D {\bf 52}, 5366 (1995); D.~Ring, S.~Urano and R.~Arnowitt,
Phys.\ Rev.\ D {\bf 52}, 6623 (1995).


\bibitem{cleo}
S. Chen et.al. (CLEO Collaboration), Phys. Rev. Lett. {\bf 87}, 251807 
(2001). 

\bibitem{belle}
H. Tajima, talk at the 20th International Symposium on 
Lepton-Photon Interactions", Rome, July 2001.

\bibitem{aleph}
R. Barate et.al., Phys. Lett. {\bf B429}, 169(1998).



\bibitem{sugra}
SUGRA Working Group Collaboration (S. Abel et. al.), arXiv:hep-ph/0003154.



\bibitem{susybsgamma1} 
P. Nath and R. Arnowitt, \Journal{\PLB}{336}{395}{1994};
\Journal{\PRL}{74}{4592}{1995};
F.~Borzumati, M.~Drees and M.~Nojiri, \Journal{\PRD}{51}{341}{1995};
H. Baer, M. Brhlik, D. Castano and  X. Tata, \Journal{\PRD}{58}
{015007}{1998}.

\bibitem{susybsgamma2}
M. Carena, D. Garcia, U. Nierste, C.E.M. Wagner, Phys. Lett. 
{\bf B499} 141 (2001); 
G. Degrassi, P. Gambino, G.F. Giudice, JHEP 0012, 009 (2000) and 
references therein; 
W. de Boer, M. Huber, A.V. Gladyshev, D.I. Kazakov, 
Eur.\ Phys.\ J.\ C {\bf 20}, 689 (2001).



\bibitem{smbsgamma}
P. Gambino and M. Misiak, Nucl. Phys. {\bf B611}, 338 (2001); 
 P. Gambino and U. Haisch,  JHEP 0110, 020 (2001). 
See  also T. Hurth, hep-ph/0106050. For previous
analysis see, A.L. Kagan and M. Neubert, Eur. Phys. J. C7, 5(1999).


\bibitem{lephiggs}
 [LEP Higgs Working Group Collaboration],
``Searches for the neutral Higgs bosons of the MSSM:
 Preliminary combined  results using LEP data collected at energies up
  to 209-GeV,'' arXiv:hep-ex/0107030.


 \bibitem{brown}
H.N. Brown et al., Muon ($g-2$) Collaboration, 
Phys. Rev. Lett. {\bf 86}, 2227 (2001).

\bibitem{lsp}
R.~Arnowitt and P.~Nath,
Phys.\ Rev.\ Lett.\  {\bf 69}, 725 (1992) 





\bibitem{superk} 
 Y. Totsuka, Talk at the SUSY2K conference at CERN, June 2000.
 

\bibitem{czar1}
A. Czarnecki and W.J. Marciano, {\it Nucl. Phys. (Proc. Suppl.)} {\bf B76}, 
245(1999).



\bibitem{hayakawa}
H. Hayakawa, T. Kinoshita and A. Sanda, Phys. Rev. Lett. {\bf 75},
790(1995); Phys. Rev. {\bf D54}, 3137(1996); M. Hayakwa and T. Kinoshita,
Phys. Rev. {\bf D57}, 465(1998).

\bibitem{bijnens}
J. Bijnens, E. Pallante and J. Prades, Phys.\ Rev.\ Lett.\ 
{\bf 75}, 1447(1995); ibid {\bf 75}, 3781(1995); 
E. Nucl. Phys. {\bf B474}, 379(1996). See also: 
Ref.~\cite{Bijnens2}.


\bibitem{knecht}
M. Knecht and A. Nyffeler, arXiv:hep-ph/0111058;
M. Knecht, A. Nyffeler, M. Perrottet and E. de Rafael, Phys.\ Rev.\ Lett.\ 
{\bf 88}, 071802 (2002). 

\bibitem{hkrevised}
M. Hayakawa and T. Kinoshita, arXiv:hep-ph/0112102.




\bibitem{Bijnens2}
J.~Bijnens, E.~Pallante and J.~Prades,
arXiv:hep-ph/0112255.




\bibitem{Blokland}
I.~Blokland, A.~Czarnecki and K.~Melnikov,
Phys.\ Rev.\ Lett.\  {\bf 88}, 071803 (2002).





\bibitem{Ramsey}
M.~Ramsey-Musolf and M.~B.~Wise,
 theory,'' arXiv:hep-ph/0201297.




\bibitem{davier}
 M. Davier and A. H\"ocker, \Journal {\PLB} {435} {427}{1998}.


\bibitem{hadronic}
For other assessments of the hadronic error see,
F.J. Yndurain, hep-ph/0102312; J.F. De Troconiz and F.J. Yndurain, 
arXiv:hep-ph/0106025;
S.~Narison, Phys.\ Lett.\ B {\bf 513}, 53 (2001);
K. Melnikov, Int. Jour. of Mod. Phys. {\bf A16}, 4591, (2001) 
[arXiv:hep-ph/0105267];
G. Cvetic, T. Lee and I. Schmidt, Phys.\ Lett.\ B {\bf 520}, 222 (2001). 
 For a review of status of the hadronic error see, 
  W.J. Marciano and B.L. Roberts, "Status of the hadronic 
  contribution to the muon $g-2$ value", arXiv:hep-ph/0105056;
  J. Prades, "The Standard Model Prediction for Muon $g-2$",
  arXiv:hep-ph/0108192 



\bibitem{msugra}
A.H. Chamseddine, R. Arnowitt and P. Nath, \Journal{\PRL}{49}
{970}{1982}; R. Barbieri, S. Ferrara and C.A. Savoy, \Journal{\PLB}
{119}{343}{1982}; L. Hall, J. Lykken, and S. Weinberg,
\Journal{\PRD}{27}{2359}{1983}: P. Nath, R. Arnowitt and A.H. Chamseddine,
\Journal{\NPB}{227}{121}{1983}. For reviews, see P. Nath, R. Arnowitt
and A.H. Chamseddine, "Applied N=1 Supergravity", world scientific,
1984; H.P. Nilles, Phys. Rep. {\bf 110}, 1(1984).




\bibitem{yuan}
  T. C. Yuan, R. Arnowitt, A. H. Chamseddine and P. Nath, 
 \Journal {\ZPC}{26}{407}{1984};
 D.A. Kosower, L.M. Krauss, N. Sakai, \Journal{\PLB}{133}{305}{1983};


\bibitem{lopez}
J.L. Lopez, D.V. Nanopoulos, X. Wang, \Journal{\PRD}
{49}{366}{1994}.

\bibitem{chatto}
U. Chattopadhyay and P. Nath, \Journal{\PRD}{53}{1648}{1996};
T. Moroi, \Journal{\PRD}{53}{6565}{1996}; M. Carena,
M. Giudice  and C.E.M. Wagner, \Journal{\PLB}{390}{234}{1997};
E. Gabrielli and U.  Sarid,  \Journal{\PRL}{79}{4752}{1997};
K.T. Mahanthappa and S. Oh, \Journal{\PRD}{62}{015012}{2000};
T. Blazek, arXiv:hep-ph/9912460;
U.Chattopadhyay , D. K. Ghosh and S. Roy, \Journal{\PRD}{62}{115001}{2000}.



\bibitem{in}
T. Ibrahim and P. Nath, Phys. Rev. {bf D61},095008(2000);
Phys. Rev. {\bf D62}, 015004(2000); arXiv:hep-ph/0107325



\bibitem{chatto2}
U.~Chattopadhyay and P.~Nath, in Ref.\cite{everett}.

\bibitem{everett}
L.~L.~Everett, G.~L.~Kane, S.~Rigolin and L.~Wang, 
Phys.\ Rev.\ Lett.\  {\bf 86}, 3484 (2001);
J.~L.~Feng and K.~T.~Matchev, Phys.\ Rev.\ Lett.\  {\bf 86}, 3480 (2001);
E.~A.~Baltz and P.~Gondolo, Phys.\ Rev.\ Lett.\  {\bf 86}, 5004 (2001);
U.~Chattopadhyay and P.~Nath,
Phys.\ Rev.\ Lett.\  {\bf 86}, 5854 (2001);
 S. Komine, T. Moroi, and M. Yamaguchi, Phys.\ Lett.\ B {\bf 506}, 93 (2001);
Phys.\ Lett.\ B {\bf 507}, 224 (2001);
J. Ellis, D.V. Nanopoulos, K. A. Olive, Phys.\ Lett.\ B {\bf 508}, 65 (2001);
R. Arnowitt, B. Dutta, B. Hu, Y. Santoso, 
Phys.\ Lett.\ B {\bf 505}, 177 (2001);
S. P. Martin, J. D. Wells, Phys.\ Rev.\ D {\bf 64}, 035003 (2001);
H. Baer, C. Balazs, J. Ferrandis, X. Tata, 
Phys.Rev.{\bf D64}: 035004, (2001); 
M. Byrne, C. Kolda,  J.E. Lennon, arXiv:hep-ph/0108122. 
For a more complete set
of references see, U.~Chattopadhyay and P.~Nath,
arXiv:hep-ph/0108250.

\bibitem{ccnyuk}
U.~Chattopadhyay, A.~Corsetti and P.~Nath,
arXiv:hep-ph/0201001; arXiv:hep-ph/0202275.

\bibitem{cms}
CMS Collaboration, Technical Proposal: CERN/LHCC 94-38(1994);
ATLAS Collaboration, Technical Proposal, CERN/LHCC 94-43(1944);
H. Baer, C-H. Chen, F. Paige and X. Tata, Phys. Rev. {\bf D52},
2746(1995); Phys. Rev. {\bf D53}, 6241(1996).


\bibitem{ny1}
P.~Nath and M.~Yamaguchi,
Phys.\ Rev.\ D {\bf 60}, 116004 (1999);
Phys.\ Rev.\ D {\bf 60}, 116006 (1999).
See also K.~Agashe, N.~G.~Deshpande and G.~H.~Wu,
Phys.\ Lett.\ B {\bf 489}, 367 (2000).
For a review see, P. Nath, arXiv:hep-ph/0011177

\bibitem{graesser}
M.~L.~Graesser,
Phys.\ Rev.\ D {\bf 61}, 074019 (2000)
[arXiv:hep-ph/9902310].




\bibitem{gravity}
C.~D.~Hoyle, U.~Schmidt, B.~R.~Heckel, E.~G.~Adelberger, J.~H.~Gundlach, D.~J.~Kapner and H.~E.~Swanson,
Phys.\ Rev.\ Lett.\  {\bf 86}, 1418 (2001)
[arXiv:hep-ph/0011014].

\bibitem{antoniadis}
I.~Antoniadis,
Phys.\ Lett.\ B {\bf 246}, 377 (1990);
I.~Antoniadis, K.~Benakli and M.~Quiros,
Phys.\ Lett.\ B {\bf 331}, 313 (1994)
[arXiv:hep-ph/9403290]; I.~Antoniadis, K.~Benakli and M.~Quiros,
Phys.\ Lett.\ B {\bf 460}, 176 (1999) 
[arXiv:hep-ph/9905311].



\bibitem{nyy}
P.~Nath, Y.~Yamada and M.~Yamaguchi,
Phys.\ Lett.\ B {\bf 466}, 100 (1999)
[arXiv:hep-ph/9905415];
T.~G.~Rizzo,
  colliders,'' Phys.\ Rev.\ D {\bf 61}, 055005 (2000)
[arXiv:hep-ph/9909232].


\bibitem{icn}
T.~Ibrahim, U.~Chattopadhyay and P.~Nath,
Phys.\ Rev.\ D {\bf 64}, 016010 (2001)
[arXiv:hep-ph/0102324].


\bibitem{inmuedm}
T. Ibrahim and P.Nath, Phys. Rev. {\bf D64}, 093002(2001);
J.L. Feng, K.T. Matchev, and Y. Shadmi, Nucl. Phys. {\bf B613}, 
366(2001). 


\bibitem{muedm}
Y.K. Semertzidis et.al., hep-ph/0012087

 \bibitem{bagger}
D. Pierce, J. Bagger, K. Matchev and R. Zhang, Nucl. Phys. {\bf B491},
3(1997); H. Baer, H. Diaz, J. Ferrandis and X. Tata, Phys. Rev.
{\bf D61}, 111701(2000).

\bibitem{deboer}
W. de Boer, M. Huber, A.V. Gladyshev, D.I. Kazakov, 
Eur.\ Phys.\ J.\ C {\bf 20}, 689 (2001);
W. de Boer, M. Huber, C. Sander, and D.I. Kazakov, arXiv:hep-ph/0106311;


\bibitem{susybtmass} L.J. Hall, R. Rattazzi and U. Sarid, Phys. Rev {\bf D50},
7048 (1994); R. Hempfling,  Phys. Rev {\bf D49}, 6168 (1994); M. Carena,
M. Olechowski, S. Pokorski and C. Wagner, Nucl. Phys. {\bf B426}, 269 (1994); 
D. Pierce {\it et. al.} of Ref.\cite{bagger}.


\bibitem{bf}
H. Baer and J. Ferrandis, Phys. Rev. Lett.{\bf 87}, 211803 (2001). 

\bibitem{bdr}
T.~Blazek, R.~Dermisek and S.~Raby,
Phys.\ Rev.\ Lett.\  {\bf 88}, 111804 (2002);
T. Blazek, R. Dermisek and S. Raby, arXiv:hep-ph/0107097;
R. Dermisek, arXiv:hep-ph/0108249;
S. Raby, arXiv:hep-ph/0110203. 

 \bibitem{ky}
 S. Komine and M. Yamaguchi, arXiv:hep-ph/0110032

\bibitem{cnbtau}
U.~Chattopadhyay and P.~Nath,
Phys.\ Rev.\ D {\bf 65}, 075009 (2002).

\bibitem{anderson}
G. Anderson, C.H. Chen, J.F. Gunion, J. Lykken, T. Moroi, and
Y. Yamada, arXiv:hep-ph/9609457; G. Anderson, H. Baer, C-H Chen and
X. Tata, Phys.\ Rev.\ D {\bf 61}, 095005 (2000).
 
 

\bibitem{chamoun}
N. Chamoun, C-S Huang, C Liu, and X-H Wu, Nucl.\ Phys.\ {\bf B624}, 81 (2002). 

\bibitem{ellisdark}
J. Ellis, astro-ph/0204059

\bibitem{detection}
R.~Arnowitt and P.~Nath,
Phys.\ Rev.\ D {\bf 60}, 044002 (1999);
A.~Corsetti and P.~Nath,
Int.\ J.\ Mod.\ Phys.\ A {\bf 15}, 905 (2000);
P. Belli, R. Bernbei, A. Bottino, F. Donato, N. Fornengo,
 D. Prosperi, and S. Scopel, Phys. Rev. {\bf D61}, 023512(2000);
J. Ellis, T. Falk, K. A. Olive, M. Srednicki, Astropart. Phys. {\bf 13},
181(2000);
M.E. Gomez, G. Lazarises, and C. Pallis, Phys. Lett. {\bf B487}, 
313(2000); 
J.L. Feng, K.T. Matchev. F. Wilczek, Phys. Lett. {\bf B482}, 388(2000);
Phys.\ Rev.\ D {\bf 63}, 045024 (2001);
M. Brhlik and G.L. Kane, hep-ph/0005158; 
R. Arnowitt, B. Dutta, and Y. Santoso, Nucl. Phys. {\bf B606}, 59(2001); 
J.~D.~Vergados,
Phys.\ Rev.\ D {\bf 63}, 063511 (2001); 
 T.~Nihei, L.~Roszkowski and R.~Ruiz de Austri,
JHEP {\bf 0203}, 031 (2002).
V.~A.~Bednyakov, H.~V.~Klapdor-Kleingrothaus and E.~Zaiti,
arXiv:hep-ph/0203108.

\bibitem{gomez} 
M.E. Gomez, G. Lazarides and C. Pallis, Phys. Rev. {\bf D61}, 
123512(2000).

\bibitem{cndark1}
A.~Corsetti and P.~Nath,
Phys.\ Rev.\ D {\bf 64}, 125010 (2001); hep-ph/0011313.

\bibitem{said}
M.~M.~Pavan, I.~I.~Strakovsky, R.~L.~Workman and R.~A.~Arndt,
arXiv:hep-ph/0111066; SAID pion-nucleon database, 
http://gwdac-phys.gwu.edu; A. Bottion, F. Donato, N. Fornengo and
S. Scopel, hep-ph/0111229.


\bibitem{shafi}
B.~Ananthanarayan, G.~Lazarides and Q.~Shafi,
Phys.\ Rev.\ D {\bf 44}, 1613 (1991).

\bibitem{ccn}
K.L. Chan, U. Chattopadhyay and P. Nath, \Journal {\PRD}{58}{096004}{1998}.

\bibitem{trilep}
P. Nath and R. Arnowitt, Mod. Phys.Lett.{\bf A2}, 331(1987);
H. Baer and X. Tata, Phys. Rev.{\bf D47}, 2739(1993);
V. Barger and C. Kao, Phys. Rev. {\bf D60}, 115015(1999).


\bibitem{dama} 
R. Belli et.al., Phys. Lett.{\bf B480},23(2000),
"Search for WIMP annual modulation signature:
results from DAMA/NAI-3 and DAMA/NAI-4 and the global combined
analysis", DAMA collaboration preprint INFN/AE-00/01, 1 February, 2000.


\bibitem{cdms}
R. Abusaidi et.al., Phys. Rev. Lett.{\bf84}, 5699(2000),
"Exclusion Limits on WIMP-Nucleon Cross-Section
from the Cryogenic Dark Matter Search", CDMS Collaboration preprint
CWRU-P5-00/UCSB-HEP-00-01 and astro-ph/0002471.

\bibitem{genius}
H.V. Klapor-Kleingrothaus, et.al., 
"GENIUS, A Supersensitive Germanium Detector System for Rare Events: 
Proposal", MPI-H-V26-1999, arXiv:hep-ph/9910205.



\bibitem{pdecay}
J. Ellis, D.V. Nanopoulos and S. Rudaz, Nucl. Phys. {\bf B202},
43(1982); P. Nath, R. Arnowitt and A.H. Chamseddine, Phys. Rev.
{\bf D32}, 2348(1985); Phys. Lett.{\bf B156}, 215(1985);
J. Hisano, H. Murayama and T. Yanagida, Nucl. Phys. {\bf B402}, 
46(1993); T. Goto and T. Nihei, Phys. Rev. {\bf D59}, 115009(1999);
V. Lucas and S. Raby, Phys. Rev. {\bf D55}, 6986(1997);
K.S. Babu, J.C. Pati and F. Wilczek, Nucl. Phys. {\bf B566}, 33(2000). 

\bibitem{dmr}
R.~Dermisek, A.~Mafi and S.~Raby,
Phys.\ Rev.\ D {\bf 63}, 035001 (2001)

\bibitem{mp} 
H.~Murayama and A.~Pierce,
Phys.\ Rev.\ D {\bf 65}, 055009 (2002)

 
\bibitem{aoki}
S. Aoki et.al., Phys. Rev. {\bf D62}, 014506(2000).  
   
  
\bibitem{altarelli}
G.~Altarelli, F.~Feruglio and I.~Masina,
JHEP {\bf 0011}, 040 (2000)

\bibitem{ns}
 P.~Nath and R.~M.~Syed,
Phys.\ Lett.\ B {\bf 506}, 68 (2001); 
Nucl.\ Phys.\ B {\bf 618}, 138 (2001).






\end{thebibliography}
\end{document}